\documentclass[10pt,conference]{IEEEtran}

\usepackage[T1]{fontenc}
\usepackage{cite}
\usepackage{textcomp}
\usepackage{wrapfig}
\usepackage{amsmath,amssymb}
\usepackage{graphicx}
\usepackage{booktabs}
\usepackage{placeins}
\usepackage{tabularx}
\usepackage{xcolor}
\usepackage{url}
\usepackage{xspace}

\usepackage[most]{tcolorbox}
\usepackage{listings}
\usepackage{tikz}
\usetikzlibrary{arrows.meta,positioning,fit,backgrounds}
\newtcolorbox{appendixcode}[2][]{%
  enhanced,
  breakable,
  colback=black!1,
  colframe=black!25,
  boxrule=0.6pt,
  arc=2pt,
  left=6pt,right=6pt,top=6pt,bottom=6pt,
  fonttitle=\bfseries,
  title={#2},
  #1
}
\newcommand{\proj}{\scalebox{0.9}{\textsf{ShadowProbe}}\xspace}
\xspaceaddexceptions{~}
\newcommand{\emailBench}{\scalebox{0.9}{\projectname{PyEmailBench}}\xspace}

\newcommand{\spfWca}{\hbox{SPF-WCA}\xspace}
\newcommand{\badger}{\hbox{Badger}\xspace}
\newcommand{\hotfuzz}{\hbox{HotFuzz}\xspace}
\newcommand{\acquirer}{\hbox{Acquirer}\xspace}
\newcommand{\wise}{\hbox{WISE}\xspace}

\newcommand{\singularity}{\hbox{Singularity}\xspace}
\newcommand{\hypofuzz}{\hbox{HypoFuzz}\xspace}
\newcommand{\jdkCommitUrl}{https://github.com/openjdk/jdk/tree/a3f9e222632d29982ef1463e6c391d5896524705}
\newcommand{\cpythonCommitUrl}{https://github.com/python/cpython/tree/5ab66a882d1b5e44ec50b25df116ab209d65863f}
\newcommand{\zigCommitUrl}{https://github.com/ziglang/zig/tree/1f43e049d0d46c14ce5d598f1b6587d2c043b82c}
\newcommand{\rustcCommitUrl}{https://github.com/rust-lang/rust/tree/61d7280f3c4c63fa24c56bdaa9a446151b5a30dc}
\newcommand{\vllmCommitUrl}{https://github.com/vllm-project/vllm/tree/8cac35ba435906fb7eb07e44fe1a8c26e8744f4e}
\newcommand{\jdkCommit}{\href{\jdkCommitUrl}{a3f9e22}\xspace}
\newcommand{\cpythonCommit}{\href{\cpythonCommitUrl}{5ab66a8}\xspace}
\newcommand{\zigCommit}{\href{\zigCommitUrl}{1f43e04}\xspace}
\newcommand{\rustcCommit}{\href{\rustcCommitUrl}{61d7280}\xspace}
\newcommand{\vllmCommit}{\href{\vllmCommitUrl}{8cac35b}\xspace}
\newcommand{\cpython}{\hbox{CPython}\xspace}
\newcommand{\jdk}{\hbox{JDK}\xspace}

\newcommand{\cLow}{Low\xspace}
\newcommand{\cPoly}{Poly\xspace}
\newcommand{\cExp}{Exp\xspace}

\newcommand{\cLowFull}{Low-order\xspace}
\newcommand{\cPolyFull}{Polynomial\xspace}
\newcommand{\cExpFull}{Exponential\xspace}

\newcommand{\defaultLLMModel}{DeepSeek-V4-Flash\xspace}
\newcommand{\defaultLLMModelShort}{DS-V4-F\xspace}
\newcommand{\gptFourOne}{GPT-4.1\xspace}
\newcommand{\deepseek}{DeepSeek-V3.2\xspace}

\newcommand{\numJdkFoundRaw}{28}
\newcommand{\numJdkConfirmedRaw}{5}
\newcommand{\numCpythonFoundRaw}{112}
\newcommand{\numCpythonConfirmedRaw}{61}
\newcommand{\numJdkFound}{\numJdkFoundRaw\xspace}
\newcommand{\numJdkConfirmed}{\numJdkConfirmedRaw\xspace}
\newcommand{\numJdkFixedRaw}{0}
\newcommand{\numJdkFixed}{\numJdkFixedRaw\xspace}
\newcommand{\numJdkFixingRaw}{2}
\newcommand{\numJdkFixing}{\numJdkFixingRaw\xspace}
\newcommand{\numCpythonFound}{\numCpythonFoundRaw\xspace}
\newcommand{\numCpythonConfirmed}{\numCpythonConfirmedRaw\xspace}
\newcommand{\numCpythonFixedRaw}{38}
\newcommand{\numCpythonFixed}{\numCpythonFixedRaw\xspace}
\newcommand{\numCpythonFixingRaw}{17}
\newcommand{\numCpythonFixing}{\numCpythonFixingRaw\xspace}
\newcommand{\numVLLMFoundRaw}{3}
\newcommand{\numVLLMFound}{\numVLLMFoundRaw\xspace}
\newcommand{\numVLLMConfirmedRaw}{3}
\newcommand{\numVLLMConfirmed}{\numVLLMConfirmedRaw\xspace}
\newcommand{\numVLLMFixedRaw}{1}
\newcommand{\numVLLMFixed}{\numVLLMFixedRaw\xspace}
\newcommand{\numVLLMFixingRaw}{1}
\newcommand{\numVLLMFixing}{\numVLLMFixingRaw\xspace}
\newcommand{\numZigFoundRaw}{11}
\newcommand{\numZigFound}{\numZigFoundRaw\xspace}
\newcommand{\numZigConfirmedRaw}{11}
\newcommand{\numZigConfirmed}{\numZigConfirmedRaw\xspace}
\newcommand{\numZigFixedRaw}{2}
\newcommand{\numZigFixed}{\numZigFixedRaw\xspace}
\newcommand{\numZigFixingRaw}{1}
\newcommand{\numZigFixing}{\numZigFixingRaw\xspace}
\newcommand{\numRustcFoundRaw}{8}
\newcommand{\numRustcFound}{\numRustcFoundRaw\xspace}
\newcommand{\numRustcConfirmedRaw}{4}
\newcommand{\numRustcConfirmed}{\numRustcConfirmedRaw\xspace}
\newcommand{\numRustcFixedRaw}{0}
\newcommand{\numRustcFixed}{\numRustcFixedRaw\xspace}
\newcommand{\numRustcFixingRaw}{1}
\newcommand{\numRustcFixing}{\numRustcFixingRaw\xspace}
\newcommand{\numRealWorldFoundTotal}{\number\numexpr\numCpythonFoundRaw+\numJdkFoundRaw+\numVLLMFoundRaw+\numZigFoundRaw+\numRustcFoundRaw\relax\xspace}
\newcommand{\numRealWorldConfirmedTotal}{\number\numexpr\numCpythonConfirmedRaw+\numJdkConfirmedRaw+\numVLLMConfirmedRaw+\numZigConfirmedRaw+\numRustcConfirmedRaw\relax\xspace}
\newcommand{\numRealWorldFixedTotal}{\number\numexpr\numCpythonFixedRaw+\numJdkFixedRaw+\numVLLMFixedRaw+\numZigFixedRaw+\numRustcFixedRaw\relax\xspace}
\newcommand{\numRealWorldFixingTotal}{\number\numexpr\numCpythonFixingRaw+\numJdkFixingRaw+\numVLLMFixingRaw+\numZigFixingRaw+\numRustcFixingRaw\relax\xspace}

\newcommand{\pythonAdapterLOC}{1,735\xspace}
\newcommand{\javaAdapterLOC}{1,218\xspace}
\newcommand{\rustAdapterLOC}{596\xspace}
\newcommand{\zigAdapterLOC}{391\xspace}

\newcommand{\Candidate}{\emph{Candidate}\xspace}
\newcommand{\Candidates}{\emph{Candidates}\xspace}

\newcommand{\acv}{ACV\xspace}
\newcommand{\acvs}{ACVs\xspace}

\newcommand{\numJDKFuncs}{178\,\text{k}\xspace}
\newcommand{\numCPythonFuncs}{15\,\text{k}\xspace}
\newcommand{\numCPythonEmailFuncs}{281\xspace}
\newcommand{\numCPythonEmailACVRaw}{69}
\newcommand{\numCPythonEmailACV}{\numCPythonEmailACVRaw\xspace}
\newcommand{\shadowComplexity}{{shadow complexity}\xspace}

\newcommand{\speedupSpf}{4.7$\times$}
\newcommand{\speedupBadger}{8.1$\times$}

\definecolor{BlindColorTolNine}{HTML}{006CD1}

\definecolor{BlindColorTolNine1}{HTML}{5d3131}

\definecolor{BlindColorTolNine1}{HTML}{FFA500}

\definecolor{DefinedOrange}{HTML}{ED5537}

\newcommand{\llms}{LLMs\xspace}
\newcommand{\llm}{LLM\xspace}
\newcommand{\dos}{DoS\xspace}
\newcommand{\api}{API\xspace}
\newcommand{\apis}{APIs\xspace}
\newcommand{\cves}{CVEs\xspace}
\newcommand{\vllm}{vLLM\xspace}
\newcommand{\zig}{Zig\xspace}
\newcommand{\rustc}{Rustc\xspace}
\newcommand{\jit}{JIT\xspace}
\newcommand{\gc}{GC\xspace}

\newcommand{\ChallengeOne}{Challenge~\circled{1}\xspace}
\newcommand{\ChallengeTwo}{Challenge~\circled{2}\xspace}

\usepackage[ruled,lined,linesnumbered,vlined]{algorithm2e}

\SetCommentSty{mycommfont}
\DontPrintSemicolon %

\usepackage{amsmath}
\usepackage{amsfonts}
\usepackage{amsthm}
\usepackage{balance}
\usepackage{enumitem}
\usepackage{graphicx}
\usepackage{listings}
\usepackage{multicol}
\usepackage{multirow}
\usepackage{subcaption}
\usepackage{url}
\usepackage{xspace}
\usepackage{xcolor}
\usepackage{tcolorbox}
\usepackage{colortbl}
\usepackage{xurl}

\usepackage{tikz}
\usepackage{pgfplots}
\pgfplotsset{compat=1.18}
\newcommand*\circled[1]{\tikz[baseline=(char.base)]{
    \node[shape=circle,draw,inner sep=0.5pt] (char) {\small#1};}}
\usetikzlibrary{shapes}
\usetikzlibrary{shapes.geometric}
\usetikzlibrary{arrows.meta, positioning}

\newcommand{\eg}{\mbox{e.g.}\xspace}

\definecolor{BlindColorTolOne}{HTML}{332288}
\definecolor{BlindColorTolTwo}{HTML}{117733} %
\definecolor{BlindColorTolThree}{HTML}{44AA99}
\definecolor{BlindColorTolFour}{HTML}{88CCEE}
\definecolor{BlindColorTolFive}{HTML}{DDCC77}
\definecolor{BlindColorTolSix}{HTML}{CC6677} %
\definecolor{BlindColorTolSeven}{HTML}{AA4499}
\definecolor{BlindColorTolEight}{HTML}{882255}

\definecolor{BlindColorWongOne}{HTML}{000000} %
\definecolor{BlindColorWongTwo}{HTML}{E69F00}
\definecolor{BlindColorWongThree}{HTML}{56B4E9}
\definecolor{BlindColorWongFour}{HTML}{009E73}
\definecolor{BlindColorWongFive}{HTML}{F0E442}
\definecolor{BlindColorWongSix}{HTML}{0072B2} %
\definecolor{BlindColorWongSeven}{HTML}{D55E00}
\definecolor{BlindColorWongEight}{HTML}{CC79A7}

\definecolor{mygreen}{HTML}{02818a}

\mathchardef\mhyphen="2D

\newcounter{FindingCounter}

\newcommand{\myparagraph}[1]{
  \vspace*{0.04cm}
  \noindent \textit{\textbf{#1.}}\quad
}

\newcommand{\lightparagraph}[1]{
  \vspace*{1px}
  \noindent\underline{#1.}
}

\newcommand{\projectname}[1]{\mbox{\textsf{#1}}\xspace} %
\SetKwData{AlgTrue}{true}
\SetKwData{AlgFalse}{false}

\usepackage[hidelinks]{hyperref}
\usepackage{cleveref} %

\Crefname{algocf}{Algorithm}{Algorithms}
\crefname{algocf}{Algorithm}{Algorithms}

\Crefname{algorithm}{Algorithm}{Algorithms}
\crefname{algorithm}{Algorithm}{Algorithms}

\crefname{appendix}{Appendix}{Appendices}
\Crefname{appendix}{Appendix}{Appendices}

\Crefname{figure}{Figure}{Figures}
\crefname{figure}{Figure}{Figures}

\crefname{listing}{Listing}{Listings}
\Crefname{listing}{Listing}{Listings}

\Crefname{table}{Table}{Tables}
\crefname{table}{Table}{Tables}

\crefname{thm}{Theorem}{Theorems}
\Crefname{thm}{Theorem}{Theorems}

\crefname{equation}{Equation}{Equations}
\Crefname{equation}{Equation}{Equations}

\crefformat{chapter}{\S~#2#1#3}
\crefmultiformat{chapter}{\S\S~#2#1#3}{ and~#2#1#3}{, #2#1#3}{, and~#2#1#3}

\crefformat{section}{\S~#2#1#3}
\crefmultiformat{section}{\S\S~#2#1#3}{ and~#2#1#3}{, #2#1#3}{, and~#2#1#3}

\newif\ifshowcomments
\newcommand{\setshowcomments}[1]{%
  \csname showcomments#1\endcsname
}
\setshowcomments{true}

\begin{document}

\title{ShadowProbe: Language-Extensible Detection of Hidden Algorithmic Complexity Vulnerabilities}
\author{
  \IEEEauthorblockN{
    Yuanmin Xie\IEEEauthorrefmark{1}\textsuperscript{1},
    Xiangfan Wu\IEEEauthorrefmark{2}\IEEEauthorrefmark{3}\textsuperscript{1},
    Wenhao Wu\IEEEauthorrefmark{4}\IEEEauthorrefmark{3},
    Lingyun Ying\IEEEauthorrefmark{3},
    Puzhuo Liu\IEEEauthorrefmark{1},
    \\
    Haipeng Qu\IEEEauthorrefmark{2},
    Zhongyuan Chen\IEEEauthorrefmark{2}\IEEEauthorrefmark{3},
    Min Zhou\IEEEauthorrefmark{1},
    Chengnian Sun\IEEEauthorrefmark{5}
  }
  \IEEEauthorblockA{
    \begin{tabular}{ccc}
      \IEEEauthorrefmark{1}Tsinghua University &
      \IEEEauthorrefmark{2}Ocean University of China &
      \IEEEauthorrefmark{3}QI-ANXIN Technology Research Institute \\
      \multicolumn{3}{c}{\IEEEauthorrefmark{4}Southeast University \quad \IEEEauthorrefmark{5}University of Waterloo}
    \end{tabular}
  }
  \IEEEauthorblockA{\small
    \texttt{\{xieym23@mails,liupz@mail,mzhou\}@tsinghua.edu.cn}\\
    \texttt{\{wuxiangfan@stu,quhaipeng\}@ouc.edu.cn, yinglingyun@qianxin.com}\\
    \texttt{\{ch4ml1nx,christychen516\}@gmail.com, cnsun@uwaterloo.ca}
  }
}

\maketitle
\footnotetext[1]{Both authors contributed equally to this research.}
\setcounter{footnote}{1}

\begin{abstract}
  \begin{abstract}
Algorithmic Complexity Vulnerabilities (\acvs) are a class of software flaws in which
adversarial or carefully crafted inputs trigger worst case execution behavior, leading to
severe performance degradation and Denial-of-Service (\dos) conditions. 
A key but underexplored
source of such vulnerabilities is \shadowComplexity, where non-trivial computational costs are
hidden inside seemingly benign standard library \apis. 
These costs are not visible at the call
site and can be systematically exploited to induce unexpected superlinear runtime behavior.
Existing \acv detection techniques primarily rely on fuzzing, symbolic execution, or hybrid
analysis. 
However, they are often specific to individual programming languages, require
substantial manual effort to construct execution harnesses, and depend on heavy runtime
instrumentation. This limits their scalability across large and diverse codebases.

In this work, we present \proj, a scalable and language extensible framework for discovering
\acvs through lightweight static analysis, automated reconstruction of execution contexts, and
Large Language Model (\llm) assisted test generation. 
Rather than relying on zero shot
\llm inference, \proj uses a structured multi stage pipeline. 
It first performs static screening
guided by signals from shadow complexity to identify candidate functions. It then reconstructs
minimal executable contexts from project level symbols, and finally synthesizes inputs with
controlled size to probe worst case execution behavior. Execution time measurements are used
for validation, and robust statistical growth inference is applied to separate true algorithmic
blowups from runtime noise, including garbage collection and \jit compilation effects.

We evaluate \proj on the \wise benchmark and show that it consistently outperforms existing
approaches in analysis efficiency. 
We further apply \proj to large scale software systems,
including \cpython, the \jdk, \zig, \rustc, and \vllm, where it uncovers many previously unknown
\acvs, a large portion of which have been confirmed and partially remediated by maintainers.
These results demonstrate the effectiveness of \proj in identifying hidden algorithmic risks
across diverse real world codebases.
\end{abstract}

\end{abstract}

\section{Introduction}
Software performance---typically characterized by time efficiency and execution predictability---is
a fundamental pillar of software quality.
While performance degradation is generally viewed as a routine engineering concern,
it escalates into a critical security threat when adversaries can intentionally exploit structural
bottlenecks to mount Denial-of-Service (\dos) attacks.
The primary vector for these attacks is Algorithmic Complexity Vulnerabilities
(\acvs)~\cite{DBLP:conf/uss/CrosbyW03,DBLP:conf/sp/CaiGJ09,DBLP:conf/ccs/PetsiosZKJ17,DBLP:conf/ccs/Liu022,DBLP:conf/sigsoft/AwadhutkarSHK19,DBLP:conf/ndss/BlairMAW0KE20},
classified by MITRE as CWE-407~\cite{mitre-cwe407}.
\acvs stem from inefficient algorithms that remain latent during normal operation but
exhibit catastrophic runtime slowdowns when triggered by specific, worst-case inputs.
Real-world incidents, such as the quadratic-time behavior in Python’s integer conversion (CVE-2020-10735)~\cite{cve-2020-10735}
and Perl's email parsing (CVE-2014-1474)~\cite{cve-2014-1474},
demonstrate how a single \acv can paralyze critical services across massive software ecosystems.

Detecting \acvs is inherently challenging
because their activation often depends on
intricate program logic,
such as nested loops, recursion, and specific data dependencies.
Prior research has explored \acv detection using static and dynamic techniques, including
fuzzing (\eg, SlowFuzz~\cite{DBLP:conf/ccs/PetsiosZKJ17}, \hotfuzz~\cite{DBLP:conf/ndss/BlairMAW0KE20},
\singularity~\cite{DBLP:conf/sigsoft/WeiCFFD18}),
symbolic execution (\eg, \wise~\cite{DBLP:conf/icse/BurnimJS09}, \spfWca~\cite{DBLP:conf/icst/LuckowKP17}),
and hybrid approaches (\eg, \badger~\cite{DBLP:conf/issta/NollerKP18},
\acquirer~\cite{DBLP:conf/ccs/Liu022}).
While effective on 
specific programs, these methods often rely on
extensive dynamic exploration or significant manual driver development.
This reliance limits their scalability to massive software ecosystems or language runtimes
that comprise thousands of diverse library \apis.

\myparagraph{Shadow Complexity}
The fundamental barrier to scaling these analyses is what we term \emph{shadow complexity}:
hidden computational costs induced by library \apis (or language built-ins)
that appear \emph{computationally trivial} at the call sites 
(\eg, string concatenation via \texttt{s+=x}) but exhibit \emph{non-trivial} internal complexity.
\acvs are particularly insidious when they arise from such hidden costs.
Because these abstractions are widely distributed and implicitly trusted,
a single \api with an unmodeled shadow runtime bottleneck
can quickly become an ecosystem-wide risk.
This opacity creates a significant blind spot in traditional security audits:
while reviewers typically track explicit control flow,
they often miss the amplified execution cost of seemingly harmless calls.
Addressing this gap requires scalable, automated detection methods capable of exposing these hidden bottlenecks.
However, building such a system surfaces two fundamental practical challenges:

\lightparagraph{\ChallengeOne: Scalable \& Context-Aware Screening}
As exhaustive dynamic validation of every function across a massive codebase is computationally infeasible,
the first challenge is identifying a high-coverage candidate set while preserving a strict validation budget.
This screening cannot rely solely on identifying explicit syntactic loops;
it must account for \shadowComplexity whose behavior depends on the underlying implementation.
For instance, a seemingly benign pattern like repeated string concatenation (\texttt{s += x}) in Python has implementation-dependent complexity.
While \cpython often optimizes this to near-linear behavior by reusing the left-hand string object,
other Python implementations such as PyPy may
not apply the same optimization, causing the identical source code to
exhibit quadratic copying behavior.
Consequently, initial screening signals must be computationally cheap yet sensitive to these hidden, implementation-dependent bottlenecks,
without relying on language-specific instrumentation or manual driver development.

\lightparagraph{\ChallengeTwo: Automated \& Robust Validation}
The second challenge is confirming the worst-case behavior of these candidates while minimizing false positives.
The system must automatically reconstruct compact executable contexts to synthesize worst-case inputs and faithfully exercise the target.
Furthermore, as illustrated by runtime-specific behavior in Python
implementations,
\acv detection cannot rely on isolated wall-clock measurements.
The analysis must robustly distinguish true algorithmic scaling issues from wall-clock noise
introduced by runtime effects, such as Garbage Collection (\gc) cycles and Just-In-Time (\jit) fluctuations.

\myparagraph{\proj}
To address these challenges,
we introduce \proj,
a scalable, language-extensible \acv detection system
that combines lightweight static screening,
execution context recovery, \llm-assisted worst-case input construction,
and measurement-based validation.
Unlike prior tools that are tied to specific languages
or require heavy manual effort to scale,
\proj leverages a staged architecture
to efficiently identify and automatically confirm \acvs across diverse library runtimes.
Specifically, \proj has the following advantages.

\lightparagraph{\circled{1} Shadow Complexity Guided Screening}
To address \ChallengeOne,
\proj performs lightweight screening guided by \shadowComplexity,
focusing on functions whose call chains may hide non-trivial costs behind
library \apis.
This stage combines lightweight syntactic and contextual checks to identify functions
whose behavior depends on input and may amplify hidden computational costs.
This allows \proj to maintain a high-coverage candidate set while
preserving a strict validation budget by filtering out functions with
trivial or constant-time behavior.

\lightparagraph{\circled{2} Language-Extensible Execution Context Reconstruction}
To address the manual overhead of \ChallengeTwo, 
\proj reconstructs compact executable contexts using project level symbol information.
This language-extensible approach eliminates the need for manual driver or
harness development, allowing the system to transition from source code
to executable tests without language-specific instrumentation.

\lightparagraph{\circled{3} Robust Validation \& Growth Inference}
Once an executable execution context is established,
\proj uses \llm to synthesize test inputs across varying sizes
to trigger worst-case execution and monitors wall-clock time.
To ensure the validity of our findings,
we apply a robust statistical inference module that filters out runtime
jitter from \gc and \jit fluctuations.
An \acv is confirmed only when empirical execution time exhibits
non-linear growth, providing language-independent evidence.

We evaluated \proj on \wise, a benchmark suite for worst-case complexity
analysis, and on five real-world projects: \cpython, \jdk, \zig, \rustc,
and \vllm~\cite{vllm_software}.
Across these projects, \proj uncovered \numRealWorldFoundTotal previously
unknown \acvs, of which \numRealWorldConfirmedTotal have been confirmed by
maintainers to date.
Among the confirmed cases, \numRealWorldFixingTotal are under active
maintainer processing and \numRealWorldFixedTotal have already been fixed.
These findings include cases in mature language runtimes and infrastructure
software, and have informed security discussions and planned refactoring of
the \cpython \texttt{email} module.

\myparagraph{Contributions}
We make the following contributions.
\begin{itemize}[leftmargin=*]
  \item

  We identify and define \shadowComplexity as a critical blind spot in current \acv detection—specifically, hidden computational costs within library \apis that elude manual auditing. We characterize the recurring code patterns that expose systems to this phenomenon.

  \item
  We design and implement \proj, 
  a staged, language-extensible, 
  automated, and scalable detection system 
  that reduces manual effort in \acv discovery 
  by combining shadow complexity guided candidate screening,
  language-extensible execution context reconstruction, 
  \llm-assisted worst-case input construction, and measurement-based growth validation.

  \item
  We demonstrate \proj's efficacy by uncovering previously unknown
  \acvs in widely deployed systems (\cpython, \jdk, \zig, \rustc, and \vllm).
  These findings have yielded confirmed fixes
  and assigned \cves, highlighting the tool's practical utility across diverse software ecosystems.

\end{itemize}

\section{Background \& Motivation}

\subsection{Problem Definition}
\label{sec:problem-definition}

Algorithmic complexity vulnerabilities (\acvs)~\cite{DBLP:conf/uss/CrosbyW03,DBLP:conf/ccs/PetsiosZKJ17,mitre-cwe407} refer to program behaviors
whose execution time grows superlinearly with input size (e.g.,
$\Omega(n^2)$ or worse) for some input family parameterized by \(n\),
potentially leading to performance degradation. 

\acvs often arise from interactions between control flow and data
structures, or from library and API calls whose internal implementations
exhibit non-trivial computational cost. Such behaviors may remain
invisible at the call site but can be amplified when exercised on large
or adversarial inputs.

This work focuses on identifying \acvs using syntactic signals and
execution-time evidence, where scalability issues are validated through
observed runtime growth trends under increasing input sizes.

\subsection{Large Language Models}
\llms trained on large-scale code corpora learn rich
syntactic and semantic patterns of source code~\cite{DBLP:journals/corr/abs-2107-03374}.
Unlike traditional program analysis methods that rely on manually designed,
language-specific rules, \llms enable data-driven and unified analysis across
programming languages.
\llms are also promising for detecting algorithmic complexity vulnerabilities,
as they can reason about loops and recursion
to identify potential performance risks.
In addition, \llm-generated function summaries can assist static analysis and
reduce false positives~\cite{DBLP:conf/sigsoft/Li0ZQ23}.
Beyond vulnerability detection, \llms
support cross-language code understanding and have demonstrated strong
potential in automated unit test generation, achieving higher coverage than
traditional methods~\cite{DBLP:journals/tse/SchaferNET24}.
Overall, \llms introduce data-driven
intelligence into program analysis and software testing.

Building on these capabilities, this work leverages \llms to analyze program
semantics related to \acvs.
\llms are used to synthesize executable, size-controlled input generators.
\proj then classifies complexity from measured execution-time trends.

\usetikzlibrary{arrows.meta, positioning}

\begin{figure}[t]
  \centering
  \footnotesize

  \begin{tikzpicture}[
    >=Latex,
    every node/.style={font=\footnotesize},
    box/.style={
      draw,
      rounded corners,
      fill=gray!10,
      inner sep=6pt,
      text width=0.95\columnwidth,
      align=left
    }
  ]

  \node[box] (code) {%
    \textbf{Shadow complexity in \texttt{expandvars} (original)}\\[-0.25em]
    {\ttfamily\footnotesize
    \def\I{\hspace*{1.2em}}%
    \def\II{\hspace*{2.4em}}%
    \setlength{\tabcolsep}{0pt}%
    \renewcommand{\arraystretch}{1.05}%
    \begin{tabular}{@{}l@{}}

    \textcolor{black!70}{\textbf{\# Repeated regex search}}\\
    \detokenize{while True:}\\
    \I\detokenize{m = search(path, i)}\\
    \I\detokenize{if not m: break}\\
    \I\detokenize{i, j = m.span(0)}\\
    \I\detokenize{name = m.group(1)}\\
    \I\detokenize{...}\\[0.15em]

    \textcolor{black!70}{\textbf{\# Full string reconstruction}}\\
    \I\detokenize{tail = path[j:]}\\
    \I\detokenize{path = path[:i] + value}\\
    \I\detokenize{path += tail}\\

    \end{tabular}}
  };

  \end{tikzpicture}

    \caption{A vulnerability from hidden library operations in
    \texttt{expandvars}.}
  \label{fig:expandvars}
\end{figure}
\subsection{Motivating Example: Shadow Complexity}

Figure~\ref{fig:expandvars} shows a representative example from
\cpython.
The function \texttt{expandvars} repeatedly searches a string for
variable references and reconstructs the string after each replacement.
Although the loop body appears to invoke only simple library operations,
each iteration may scan or copy a string whose length is proportional
to the input size.
For an input containing many variable references, the hidden costs in
each iteration accumulate across the loop, causing the total running
time of \texttt{expandvars} to grow quadratically.

\begin{figure*}[ht]
	\centering
	\includegraphics[width=1\textwidth]{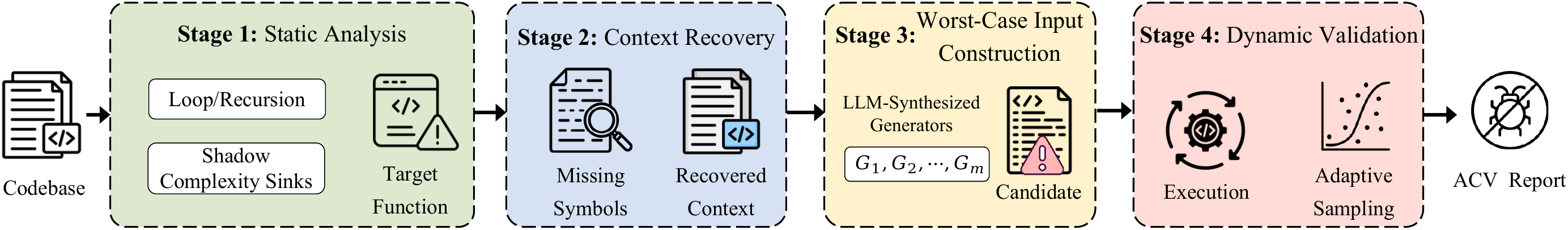}
	\caption{Overview of the \proj workflow.}
	\label{fig:workflow}
\end{figure*}

This example illustrates what we call \emph{shadow complexity}:
computational costs hidden inside \api calls or operations provided by
the language that are not visible at the call site.
Such costs often reside in library or \api implementations that appear
to be constant time from the caller's perspective, but may perform
linear or superlinear work over input data.
They are difficult to recognize when the implementation is hidden behind
library abstractions or native interfaces such as a C ABI, where the
callee's performance behavior is not explicit and may differ across
runtimes.
When such operations are invoked repeatedly under control flow that
depends on input, these hidden costs can be amplified into superlinear
runtime behavior.

This motivates \proj to consider not only explicit loops and recursion,
but also \api calls whose hidden costs can be amplified when repeatedly
executed under control flow that depends on input.

\section{Approach}

Figure~\ref{fig:workflow} illustrates the workflow of \proj.
When analyzing large-scale projects, \proj follows a four-stage pipeline that
addresses the two challenges outlined in the introduction:
the first stage performs high-density selection of target functions,
while the latter three stages jointly recover missing context,
construct verifiable worst-case inputs, and robustly infer complexity.

\noindent{\textbf{Static Analysis.}}\quad
Given the project, \proj performs lightweight static analysis to quickly
identify functions that may exhibit \acvs.

\noindent{\textbf{Context Recovery.}}\quad
To construct execution context for a target function, \proj extracts a
whole-program symbol index that records definitions, source ranges, and
reference links. 
Given a target function, \proj resolves its dependencies
via the index and expands a dependency closure over project-defined symbols.
The resulting context is rendered into a compilable artifact, which is used
by later stages to synthesize and validate a runnable \Candidate.

\noindent{\textbf{Worst-Case Input Construction.}}\quad
Given the recovered context, \proj constructs an executable \Candidate
consisting of a size-parameterized input generator and an execution harness.
The harness initializes required state, invokes the target function once,
and records execution time.
When multiple valid generators are produced, \proj keeps the one whose
probe executions exhibit the strongest empirical growth.

\noindent{\textbf{Dynamic Validation.}}\quad
\proj validates each \Candidate by executing it under 
increasing input sizes. 
Execution times are measured for each input size, 
and the resulting growth trend is used to infer the empirical complexity of the target function.

\subsection{Static Analysis}
This stage addresses \ChallengeOne by extracting a focused set of
\emph{target functions} from a given project through a lightweight,
language-extensible static scan.
For example, \cpython contains more than \numCPythonFuncs functions,
while the \jdk contains more than \numJDKFuncs functions.
\footnote{
    These numbers are obtained by counting function and method definitions 
    using \texttt{tree-sitter}~\cite{treesitter}.}
However, only a small fraction of these functions can actually trigger an
\acv, making exhaustive complexity analysis both inefficient and unnecessary.
To narrow down the search space, we apply this scan using code patterns that
are known to be prone to superlinear time complexity.
Among such patterns, a natural starting point is explicit loop and recursion
structures, as they are common syntactic indicators of potentially
superlinear behavior.
However, relying solely on these structures remains too coarse at
large-project scale.
We therefore incorporate \emph{shadow complexity sinks} to refine the static
patterns and improve the density of selected target functions.

Specifically, we define two categories of code patterns that are prone to
triggering \acvs.

\myparagraph{Nested Loops and Recursion}
Nested loops and recursion are common structural features of \acvs: when
iteration bounds or recursion depth depend on external inputs, runtime costs
can be multiplicatively amplified.
Prior work (e.g., DISCOVER~\cite{DBLP:conf/sigsoft/AwadhutkarSHK19} and
\acquirer~\cite{DBLP:conf/ccs/Liu022}) likewise often takes loops as an entry
point, first locating input-sensitive loop structures and then applying manual
inspection or dynamic validation to confirm risk.
Motivated by this, we leverage loop/recursion structure to identify and
prioritize candidate \acvs.

\myparagraph{Shadow Complexity Sinks}
\label{par:shadow-complexity-sinks}
Shadow complexity sinks are library APIs whose
cost is implicit at the call site but depends on the size of an input-derived
operand.
When invoked inside input-dependent loops or recursion, such hidden costs can be
repeatedly amplified and may trigger \acvs.
Table~\ref{tab:static-analysis-representative-calls}
lists representative sink families across common language runtimes.
\proj uses these families to refine screening based on loops and
recursion.

Figure~\ref{fig:str-slicing-complexity} shows a representative 
Python example where $N=|\texttt{methodname}|$ 
and the worst-case input is $(\texttt{<})^N$.
The guard slice \texttt{methodname[:1]} copies at most one character.
By contrast, because \cpython strings are immutable,
\texttt{methodname[1:]} creates a new string by copying the remaining suffix.
When the current length is $k$, this slice costs $\Theta(k)$, so the total
work is $\sum_{k=1}^{N}\Theta(k)=\Theta(N^2)$.
This copying cost is hidden in slicing and grows with string length.
\proj uses such hidden costs together with loops controlled by input to
select target functions.

\begin{figure}[htb]
\centering

\definecolor{YHL}{HTML}{FFF2CC}
\definecolor{RHL}{HTML}{FFDCD7}
\newcommand{\yhl}[1]{%
  \tikz[baseline=(t.base)]\node[
    fill=YHL, rounded corners=1pt,
    inner sep=0.6pt, outer sep=0pt
  ] (t) {\strut #1};%
}
\newcommand{\rhl}[1]{%
  \tikz[baseline=(t.base)]\node[
    fill=RHL, rounded corners=1pt,
    inner sep=0.6pt, outer sep=0pt
  ] (t) {\strut #1};%
}

\def\NodeDistV{1mm}          %
\def\NodeDistH{8mm}          %
\def\CodeLineSep{0pt}      %
\def\CodeBoxInnerSep{0pt}  %
\def\TopToCostShift{-3mm}    %
\def\CostTitleToBarsShift{-7mm} %
\def\BarH{0.28}              %
\def\BarDy{0.50}             %
\def\EllipsisExtra{0.05}     %
\def\SumDyFactor{4.8}        %

\resizebox{\linewidth}{!}{%
\begin{tikzpicture}[
  font=\small,
  >=Latex,
  node distance=\NodeDistV and \NodeDistH
]

\node[anchor=west, font=\ttfamily\small] (l1) at (0,0)
  {\yhl{while} methodname[:1] == '\textless':};

\node[anchor=west, font=\ttfamily\small, below=\CodeLineSep of l1] (l2)
  {\hspace*{2em}methodname = methodname\rhl{[1:]};};

\begin{scope}[on background layer]
  \node[draw, rounded corners, fill=gray!10, inner sep=\CodeBoxInnerSep, fit=(l1)(l2)] (codebox) {};
\end{scope}

\node[
  draw, rounded corners, fill=YHL,
  align=center,
  text width=2.6cm,
  inner sep=3pt,
  right=4mm of codebox.north east, anchor=north west
] (c1)
{LOOP};

\node[
  draw, rounded corners, fill=RHL,
  align=center,
  text width=2.6cm,
  inner sep=3pt,
  below=2mm of c1
] (c2)
{Shadow};

\draw[->] (c1.west) -- (l1.east);
\draw[->] (c2.west) -- (l2.east);

\node[fit=(codebox)(c1)(c2), inner sep=0pt] (topfit) {};

\coordinate (cost0) at ([yshift=\TopToCostShift]topfit.south west);

\node[anchor=west] at (cost0)
{\textbf{Worst case:} \texttt{"<"} repeated $N$ times $\Rightarrow$ $N$ slices};

\def\N{20}
\def\sx{0.24}
\def\h{\BarH}
\def\dy{\BarDy}

\pgfmathtruncatemacro{\NmOne}{\N-1}
\pgfmathtruncatemacro{\NmTwo}{\N-2}
\pgfmathsetmacro{\Wmax}{\N*\sx}
\pgfmathsetmacro{\Wone}{\NmOne*\sx}
\pgfmathsetmacro{\Wtwo}{\NmTwo*\sx}
\pgfmathsetmacro{\Xlab}{\Wmax+0.8}
\pgfmathsetmacro{\Whalf}{0.5*\Wmax}

\pgfmathsetmacro{\EllipsisY}{-3*\dy+\EllipsisExtra}
\pgfmathsetmacro{\SumY}{-\SumDyFactor*\dy}

\begin{scope}[shift={(cost0)}, yshift=\CostTitleToBarsShift]

  \draw[fill=gray!15] (0, 0*\dy) rectangle (\Wmax, 0*\dy+\h);
  \foreach \i in {1,...,\N} {
    \node[font=\ttfamily\scriptsize] at ({(\i-0.5)*\sx}, {0*\dy+0.5*\h}) {\textless};
  }
  \node[anchor=west] at (\Xlab, {0*\dy+0.5*\h}) {$\Theta(N)$};

  \draw[fill=gray!15] (0,-1*\dy) rectangle (\Wone,-1*\dy+\h);
  \foreach \i in {1,...,\NmOne} {
    \node[font=\ttfamily\scriptsize] at ({(\i-0.5)*\sx}, {-1*\dy+0.5*\h}) {\textless};
  }
  \node[anchor=west] at (\Xlab, {-1*\dy+0.5*\h}) {$\Theta(N-1)$};

  \draw[fill=gray!15] (0,-2*\dy) rectangle (\Wtwo,-2*\dy+\h);
  \foreach \i in {1,...,\NmTwo} {
    \node[font=\ttfamily\scriptsize] at ({(\i-0.5)*\sx}, {-2*\dy+0.5*\h}) {\textless};
  }
  \node[anchor=west] at (\Xlab, {-2*\dy+0.5*\h}) {$\Theta(N-2)$};

  \node at (\Whalf, \EllipsisY) {$\vdots$};

  \draw[fill=gray!15] (0,-4*\dy) rectangle (\sx,-4*\dy+\h);
  \node[font=\ttfamily\scriptsize] at ({0.5*\sx}, {-4*\dy+0.5*\h}) {\textless};
  \node[anchor=west] at (\Xlab, {-4*\dy+0.5*\h}) {$\Theta(1)$};

  \node[anchor=west,align=left] at (0,\SumY)
  {$T(N)=\Theta\!\left(\frac{N(N+1)}{2}\right)=\Theta(N^2)$};

\end{scope}

\end{tikzpicture}
}

\caption{The $\Theta(N^2)$  case in \texttt{ CPython/.../editor.py}.}
\label{fig:str-slicing-complexity}
\end{figure}

\begin{table}[t]
\centering
\caption{Representative shadow-complexity sink families.}
\label{tab:static-analysis-representative-calls}
\footnotesize
\setlength{\tabcolsep}{3pt}
\begin{tabularx}{\linewidth}{@{}l >{\raggedright\arraybackslash}X >{\raggedright\arraybackslash}X@{}}
\toprule
Family & Representative operations & Hidden cost source \\
\midrule
Text processing &
\texttt{substr}, \texttt{slicing}, \texttt{split},
\texttt{join} &
copying, scanning, or rebuilding strings \\
Collections &
\texttt{contains}, \texttt{index}, \texttt{sort}, \texttt{copyOf} &
linear scans, copying, or reordering \\
Regular expressions &
\texttt{find}, \texttt{search}, \texttt{matches}, \texttt{sub} &
pattern matching over variable-length input \\
Parsing and serialization &
\texttt{parse}, \texttt{readValue}, \texttt{decode},
\texttt{format} &
recursive parsing or object reconstruction \\
Implicit iteration &
\texttt{map}, \texttt{filter}, \texttt{sum}, \texttt{min},
\texttt{items} &
callbacks or iteration hidden behind APIs \\
\bottomrule
\end{tabularx}
\end{table}

\subsection{Context Recovery}
As a prerequisite for addressing \ChallengeTwo,
\proj recovers the context needed to construct worst-case inputs.
For a target function, the function body alone is often insufficient:
worst-case behavior may depend on surrounding project context.
Without this context, input construction may fail to produce executable tests
or may exercise only partial paths.
We therefore recover a compact context before constructing worst-case inputs.

\proj performs context recovery by first building a symbol index for the
project.
The index records program symbols, their source ranges, 
and reference information extracted from the project source code.
Starting from the target function, \proj expands a bounded dependency closure:
when a recovered symbol references another symbol defined in the project, the
corresponding definition is retrieved from the index and added to the context.
The closure also preserves enclosing declarations when they are required to make
the recovered fragment meaningful.
This process produces a compact context containing the target and the
dependencies needed by later stages, while avoiding the cost of
including the entire project.

The recovered context is emitted as a context artifact and rendered into a
compilable code fragment, which serves as the foundation for subsequent
worst-case input construction.
The context artifact records recovered symbols and unresolved references, and
the rendered fragment provides the source-level definitions used by the input
construction stage.

Figure~\ref{fig:ctx-recovery} illustrates the context recovery process for a
representative target function, \texttt{treeSearch}.
The process begins from an unresolved context in which the target function
references symbols whose definitions are not yet available.
The recovery procedure resolves these references through the symbol index and
incorporates the corresponding definitions into the context.
The process continues until the dependency closure reaches a fixed point or a
predefined recovery budget is exhausted.
This recovery procedure enables subsequent analysis of the target function under
an execution context with resolved dependencies, without requiring the full
project source code.

\usetikzlibrary{arrows.meta, positioning}

\begin{figure}[t]
  \centering
  \footnotesize

  \begin{tikzpicture}[
    >=Latex,
    node distance=4mm,
    every node/.style={font=\footnotesize},
    box/.style={
      draw,
      rounded corners,
      fill=gray!8,
      inner sep=4pt,
      text width=0.93\columnwidth,
      align=left
    },
    arr/.style={->, thick}
  ]

  \node[box] (target) {%
    \textbf{Target function}\\[-0.2em]
    {\ttfamily
    TreeNode treeSearch(TreeNode x, int k) \{ ... \}}\\[0.2em]
    {\color{red!70!black}\textit{Unresolved symbol:} \texttt{TreeNode}}
  };

  \node[box, below=of target] (trace) {%
    \textbf{Iterative context recovery}\\[-0.2em]
    {\ttfamily
    treeSearch
    $\;\rightarrow\;$ TreeNode
    $\;\rightarrow\;$ Node}\\[0.3em]
    \begin{tabular}{@{}ll@{}}
      \texttt{TreeNode} & required by parameter/return types\\
      \texttt{Node}     & required by \texttt{TreeNode}'s superclass relation
    \end{tabular}
  };

  \node[box, below=of trace] (resolved) {%
    \textbf{Recovered executable context}\\[-0.2em]
    {\ttfamily
    class TreeNode extends Node \{ ... \}\\
    class Node \{ int key() \{ ... \} \}}\\[0.2em]
    {\color{green!45!black}\textit{Resolved}}
  };

  \draw[arr] (target) -- (trace);
  \draw[arr] (trace) -- (resolved);

  \end{tikzpicture}

  \caption{Context recovery for \texttt{treeSearch}.}
  \label{fig:ctx-recovery}
\end{figure}

\subsection{Worst-Case Input Construction}

Given the recovered context, this stage constructs an executable
\Candidate that can be used for subsequent dynamic validation.
A \Candidate consists of a size-parameterized input generator and an
execution harness.
Given an input size~$n$, the generator produces concrete arguments and
optional state-building inputs intended to exercise high-cost execution
paths of the target function.

Formally, the generator is a size-parameterized mapping
\[
G: \mathbb{N} \rightarrow A_1 \times \cdots \times A_k,
\]
where $G(n) = (arg_1, \ldots, arg_k)$ denotes the inputs consumed by the
execution harness.
For simple targets, these arguments are passed directly to the target
function.
For targets whose behavior depends on receiver objects or internal state,
they may instead include constructor arguments or state-building parameters.
The parameter~$n$ controls the dominant size measure of the generated inputs,
which is required to grow as~$\Theta(n)$.

\begin{figure}[t]
  \centering

\begin{tikzpicture}[
  font=\rmfamily\scriptsize,
  title/.style={font=\rmfamily\bfseries\small},
  card/.style={draw=black!45, fill=gray!4, rounded corners=3pt, line width=0.45pt},
  row/.style={draw=black!20, fill=white, rounded corners=2pt, line width=0.35pt, inner sep=3pt, align=left, minimum height=0.48cm},
  tag/.style={font=\rmfamily\bfseries\scriptsize, text=black!88}
]

\node[title, anchor=north west] (prompttitle) at (0,0) {Prompt};

\newcommand{\promptrow}[2]{%
  \begin{tabular}{@{}p{1.05cm}@{\hspace{0.12cm}}>{\raggedright\arraybackslash}p{5.72cm}@{}}
    \textbf{#1} & #2
  \end{tabular}%
}

\node[row, text width=7.22cm, anchor=north west] (r1) at ([yshift=-0.12cm]prompttitle.south west)
{\promptrow{Input}{target function; recovered context; executable Candidate harness}};
\node[row, text width=7.22cm, below=0.06cm of r1] (r2)
{\promptrow{Task}{generate three independent \texttt{gen\_inputs(n)} strategies}};
\node[row, text width=7.22cm, below=0.06cm of r2] (r3)
{\promptrow{Contract}{replace only the input generator; preserve the target call}};
\node[row, text width=7.22cm, below=0.06cm of r3] (r4)
{\promptrow{Guards}{valid $\Theta(n)$ inputs; $O(n)$ setup; no target calls or artificial work}};
\node[row, text width=7.22cm, below=0.06cm of r4] (r5)
{\promptrow{Output}{list of \{name, generator\_source\} strategies}};

\begin{scope}[on background layer]
  \node[card, fit=(prompttitle)(r1)(r2)(r3)(r4)(r5), inner sep=0.16cm] (promptcard) {};
\end{scope}

\end{tikzpicture}

  \caption{Prompt structure for synthesizing candidate worst-case input generators.
  }
  \label{fig:worst-case-input-prompt}
\end{figure}
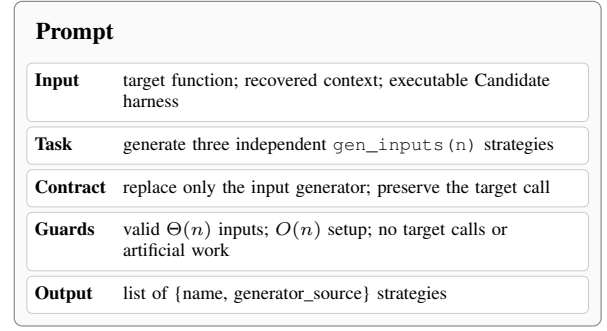

The generator is synthesized by an \llm rather than manually designed.
The prompt provides the target signature, recovered context, harness
contract, and input-size parameter~$n$.
Figure~\ref{fig:worst-case-input-prompt} summarizes 
this prompt structure.
The \llm is asked to return executable generator code that maps~$n$ to
inputs for the harness.
Compilation errors, execution failures, and size-check violations are used
as repair feedback.

Before accepting a generator, \proj performs a lightweight size check over a
small range of input sizes.
Generators whose produced inputs do not preserve the intended size relation
are rejected and repaired.
Since a single generator may miss the intended worst-case path, \proj asks
the \llm to propose multiple input-generation strategies under the same
harness.
Each strategy describes a different way to construct inputs whose size is
controlled by~$n$ and whose structure is intended to exercise a high-cost
execution path.
The strategies are instantiated as candidate generators, and \proj keeps the
valid generator with the strongest measured growth on probe executions.

Formally, given a set of candidate generators
$\mathcal{G}={G_1,\ldots,G_m}$, \proj selects
\[
G^* = \arg\max_{G_i \in \mathcal{G}} \mathrm{score}(G_i),
\]
where $\mathrm{score}(G_i)$ is computed from the measured growth of the
target execution time on probe input sizes.
The selected generator~$G^*$ is embedded with the harness to form the final
\Candidate.
Figure~\ref{fig:candidate-onebox} illustrates a synthesized \Candidate for
\texttt{\_parseparam} in \cpython.

\usetikzlibrary{arrows.meta, positioning}

\begin{figure}[t]
  \centering
  \footnotesize

  \begin{tikzpicture}[
    >=Latex,
    every node/.style={font=\footnotesize},
    box/.style={
      draw,
      rounded corners,
      fill=gray!10,
      inner sep=6pt,
      text width=0.95\columnwidth,
      align=left
    }
  ]

  \node[box] (cand) {%
    \textbf{Example synthesized \Candidate (\texttt{\_parseparam})}\\[-0.25em]
    {\ttfamily\footnotesize
    \def\I{\hspace*{1.2em}}%
    \def\II{\hspace*{2.4em}}%
    \setlength{\tabcolsep}{0pt}%
    \renewcommand{\arraystretch}{1.05}%
    \begin{tabular}{@{}l@{}}

    \textcolor{black!70}{\textbf{\# Context (abbrev.)}}\\
    \detokenize{import os, time}\\
    \textcolor{black!45}{\texttt{\# other stubs/helpers omitted}}\\

    \textcolor{black!70}{\textbf{\# Target: \_parseparam}}\\
    \detokenize{def _parseparam(s):}\\
    \I\detokenize{while s[:1] == ';':}\\
    \II\detokenize{s = s[1:]}\\
    \II\detokenize{end = s.find(';')}\\
    \II\detokenize{...}\\
    \I\detokenize{return plist}\\

    \textcolor{black!70}{\textbf{\# Generator $G(n)$ (size-controlled)}}\\
    \detokenize{def gen_inputs(n):}\\
    \I\detokenize{header = 'text/plain;charset="' + ('\\"x;' * n) + '"'}\\
    \I\detokenize{return (header,)}\\

    \textcolor{black!70}{\textbf{\# Main}}\\
    \detokenize{if __name__ == "__main__":}\\
    \I\detokenize{n = int(os.environ.get("HARNESS_N","100"))}\\
    \I\detokenize{args = gen_inputs(n)}\\
    \I\detokenize{t0 = time.perf_counter()}\\
    \I\detokenize{_parseparam(*args)}\\
    \I\detokenize{t1 = time.perf_counter()}\\
    \I\detokenize{print(f"TARGET_TIME_SECONDS={t1-t0}")}\\

    \end{tabular}}
  };

  \end{tikzpicture}

  \caption{A \Candidate for \texttt{\_parseparam}.}
  \label{fig:candidate-onebox}
\end{figure}

\subsection{Dynamic Validation}

To make worst-case claims verifiable and to robustly determine
complexity (\ChallengeTwo), \proj grounds its judgment in measured
scaling behavior rather than direct \llm inference.
Given an accepted \Candidate, \proj executes it with increasing input
sizes, collects runtime measurements $(n,t)$, and performs
curve-fitting-based growth inference over the measured trace.
This measurement-based validation reduces the risk that inaccurate
\llm reasoning is directly reported as an \acv.
As a design principle, \proj avoids instrumentation of the program under
test to preserve generality.
Instead, it measures the execution time of the target function in real
execution scenarios, making validation applicable across diverse runtimes.

\proj adopts adaptive sampling to characterize execution-time growth under
noisy wall-clock measurements.
Because wall-clock measurements may be affected by runtime effects such as
\gc and \jit, timings on very small inputs can be less reliable, whereas
overly large inputs are more likely to time out.
\proj therefore first performs boundary probing to identify a feasible
input-size range.
The lower bound avoids trivial executions with less reliable timings, while
the upper bound limits excessive timeouts.
Algorithm~\ref{alg:adaptive-sampling} illustrates the sampling procedure.

Within the identified range, \proj prioritizes logarithmically uniform input
sizes.
This choice follows the common model $T(n)=Cn^k$, where the complexity
order is reflected by the slope in log--log space.
Linear spacing would allocate many samples to the low-$n$ region, where
execution times are often less informative for fitting growth trends.
When the interval becomes too narrow for logarithmic subdivision, \proj
falls back to the arithmetic mean to ensure progressive sampling.

\begin{algorithm}[t]
\caption{Adaptive Logarithmic Sampling.}
\label{alg:adaptive-sampling}
\small
\KwIn{\Candidate $C$, range $[n_{\min}, n_{\max}]$,
budget $B$, max samples $k$}
\KwOut{sampled points $\mathcal{D}$}

$\mathcal{D} \leftarrow
\{(n_{\min}, t_C(n_{\min})), (n_{\max}, t_C(n_{\max}))\}$\;
$\mathcal{Q} \leftarrow \{(n_{\min}, n_{\max})\}$\;

\While{$\mathcal{Q} \neq \emptyset$ and $|\mathcal{D}| < k$
	and $B$ not exhausted}{
	Pop $(n_{\ell}, n_{h})$ from $\mathcal{Q}$\;
	\lIf{$n_{h} \le n_{\ell} + 1$}{\textbf{continue}}

	$n_{m} \leftarrow \lfloor \sqrt{n_{\ell} n_{h}} \rfloor$\;
	\If{$n_{m} \notin (n_{\ell}, n_{h})$}{
		$n_{m} \leftarrow \lfloor (n_{\ell} + n_{h}) / 2 \rfloor$\;
	}
	\lIf{$n_{m} \notin (n_{\ell}, n_{h})$}{\textbf{continue}}

	Run $C$ with input size $n_m$ to obtain $t_C(n_m)$\;
	\If{the run succeeds}{
		$\mathcal{D} \leftarrow
		\mathcal{D} \cup \{(n_m, t_C(n_m))\}$\;
		$\mathcal{Q} \leftarrow
		\mathcal{Q} \cup \{(n_{\ell}, n_m), (n_m, n_h)\}$\;
	}
}
\Return{$\mathcal{D}$}

\end{algorithm}

For each sampled input size, \proj executes the \Candidate three
times and aggregates the resulting measurements into a representative
runtime for that size.
This repeated measurement also reduces the influence of transient runtime noise.
Executions that fail or exceed the per-run timeout are discarded.
After collecting sampled data points, \proj classifies the measured growth
into four categories: \emph{\cLowFull} (at most $O(n \log n)$), 
\emph{\cPolyFull} (e.g., $O(n^2)$,
$O(n^3)$), \emph{\cExpFull} (e.g., $O(2^n)$, $O(n!)$), and
\emph{Unknown}.
\emph{Unknown} is assigned when valid data are unavailable or insufficient for reliable classification.
This coarse-grained taxonomy suffices to separate low-order behavior from
\acv-relevant polynomial or exponential growth.

\begin{table*}[t]
  \centering
  \caption{Analysis results on the \wise benchmark across different tools.}
  \label{tab:combined-results}
  \setlength{\tabcolsep}{5pt}
  \renewcommand{\arraystretch}{1.05}
  \small

  \begin{tabular}{@{}l c ccc ccc ccc@{}}
    \toprule
    \multirow{2}{*}{Benchmark}
      & \multirow{2}{*}{Real}
      & \multicolumn{3}{c}{\textbf{\proj}}
      & \multicolumn{3}{c}{\spfWca}
      & \multicolumn{3}{c}{\badger} \\
    \cmidrule(lr){3-5}\cmidrule(lr){6-8}\cmidrule(lr){9-11}
      &
      & Comp. & Exec. & Range
      & Comp. & Exec. & Range
      & Comp. & Exec. & Range \\
    \midrule

    BellmanFord
      & \cPoly
      & \cPoly & 43 & [1,\,19296]
      & \cPoly & 12 & [1,\,12]
      & NA     & 5  & [2,\,6] \\

    BinaryTree.search
      & \cLow
      & \cLow & 29 & [1,\,54288]
      & \cLow & 15 & [1,\,15]
      & NA    & 5  & [2,\,6] \\

    HeapInsert
      & \cLow
      & \cLow & 43 & [1,\,268435453]
      & \cLow & 100 & [1,\,100]
      & NA    & 5   & [2,\,6] \\

    RedBlackTree.insert
      & \cLow
      & \cLow & 73 & [1,\,346721480]
      & \cLow & 120 & [1,\,120]
      & NA    & 5   & [2,\,6] \\

    Dijkstra
      & \cPoly
      & \cPoly & 54 & [1,\,77432]
      & \cPoly & 30 & [1,\,30]
      & NA     & 5  & [2,\,6] \\

    SortedListInsert.insert
      & \cLow
      & \cLow & 74 & [1,\,551857236]
      & \cLow & 20 & [1,\,20]
      & NA    & 5  & [2,\,6] \\

    TSP
      & \cExp
      & \cExp & 12 & [1,\,22]
      & \textcolor{red}{\cPoly} & 7 & [1,\,7]
      & NA    & 5 & [2,\,6] \\

    QuickSort
      & \cPoly
      & \cPoly & 81 & [1,\,723148842]
      & \cPoly & 20 & [1,\,20]
      & NA     & 5  & [2,\,6] \\

    MergeSort
      & \cLow
      & \cLow & 75 & [1,\,47163520]
      & \cLow & 50 & [1,\,50]
      & NA    & 5  & [2,\,6] \\

    InsertionSort
      & \cPoly
      & \cPoly & 59 & [1,\,337098]
      & \cPoly & 6  & [1,\,6]
      & NA     & 5  & [2,\,6] \\

    \midrule
    Avg.\ Time (s)
      &
      & \multicolumn{3}{c}{379}
      & \multicolumn{3}{c}{1790}
      & \multicolumn{3}{c}{3085} \\

    \bottomrule
  \end{tabular}

  \vspace{2pt}
  \footnotesize{
    \textbf{Legend:}
    \cLow = \cLowFull;
    \cPoly = \cPolyFull;
    \cExp = \cExpFull;
    NA = not applicable.
  }
\end{table*}

\section{Evaluation}

\proj is implemented in Rust and follows a language-extensible design.
It leverages \llms to reason about program semantics and execution behavior,
without relying on language-specific analyses.
In our evaluation, we instantiate \proj for four widely used programming
languages, Java, Python, Zig, and Rust, to demonstrate its effectiveness in practical
settings.
We evaluate our approach by answering the following research questions:

\begin{description}[topsep=0pt, leftmargin=*]
	\item [RQ1:]
	      How effective is \proj at identifying algorithmic complexity issues compared to existing tools?
	\item [RQ2:]
	      What are the contributions of \proj's components, and how does the choice of
	      LLM model affect its performance?
	\item [RQ3:]
	      Can \proj detect \acvs in real-world software projects?
\end{description}

\subsection{Experimental Setup}
\label{sec:experimental-setup}
To address \textbf{RQ1}, we compare \proj with two representative and publicly
available \acv analysis tools, namely \spfWca~\cite{DBLP:conf/icst/LuckowKP17} and
\badger~\cite{DBLP:conf/issta/NollerKP18}. \spfWca infers algorithmic complexity
by symbolic execution over worst-case paths, which is conceptually closest to
our approach, while \badger combines fuzzing and concolic execution to search
for worst-case inputs under fixed input sizes.
Other \acv detectors, such as \hotfuzz~\cite{DBLP:conf/ndss/BlairMAW0KE20} and
\acquirer~\cite{DBLP:conf/ccs/Liu022}, are not publicly available and are thus
excluded from comparison.
Since both \spfWca and \badger target Java programs, we use the Java-based
\wise~\cite{DBLP:conf/icse/BurnimJS09} benchmark provided in the \spfWca
repository.
\proj can analyze the entire \wise benchmark automatically, whereas \badger
requires significant manual effort.

To address \textbf{RQ2}, we use the \emailBench benchmark, derived from the
\texttt{email} module of \cpython, a widely used module from the Python
standard library, to evaluate the contributions of \proj's major
components as well as the sensitivity of its performance to the choice
of underlying \llm.
Static analysis yields \numCPythonEmailFuncs target functions.
Based on the constructed ground truth, \numCPythonEmailACV of them exhibit
\acv behavior.
We treat analysis results classified 
as \cPolyFull or \cExpFull as indicating the presence of an \acv. 
Ground truth is constructed by systematically inspecting all target
functions in the \texttt{email} module, followed by independent manual review
by graduate student security researchers with more than three years of
security experience. Any disagreements are resolved through discussion to
reach a final consensus.
We release this set as the \emailBench benchmark for evaluating
complexity-analysis techniques.

All experiments were conducted on an AMD machine with 32 threads and
96\,GB of memory.
\defaultLLMModel was used as the underlying \llm in all experiments,
except for those in Section~\ref{sec:sensitivity-analysis}.
We additionally evaluate \gptFourOne~\cite{openai2025gpt41} and
\deepseek~\cite{deepseekai2025deepseekv32} to 
compare the impact of different \llms.
This comparison examines whether \proj remains effective across different
\llm backends under the same analysis pipeline.

\subsection{Benchmark Analysis}
Functions from the \wise benchmark are analyzed, and the results are evaluated
from two complementary perspectives:
(1) the agreement between the time complexity reported by each tool and the
ground-truth complexity, and
(2) the average analysis time of the different tools
(Table~\ref{tab:combined-results}).

Table~\ref{tab:combined-results} summarizes the analysis results for the \wise
benchmark across different tools.
The column \emph{Real} reports the ground-truth worst-case time complexity for
each benchmark.
For \proj and \spfWca, the table reports the inferred complexity class together
with the number of executions performed and the corresponding input size
ranges.
\badger does not infer time complexity and is therefore reported as NA in the
complexity column, while its execution counts and input size ranges are still
shown for comparison.

With dynamic validation, \proj correctly identifies the ground-truth
worst-case complexity class for all benchmarks in the \wise dataset.
This result shows that \proj can expose worst-case time-complexity behavior
across both graph algorithms and data-structure operations.
In contrast, \spfWca underestimates \textsc{TSP}, reporting \cPolyFull rather
than the ground-truth \cExpFull class, which is consistent with the results
reported in its original paper.
For the remaining benchmarks, \spfWca agrees with the ground-truth complexity
class reported in the \wise benchmark.

Table~\ref{tab:combined-results} also reports the average analysis time of \proj,
\spfWca, and \badger on the \wise dataset, 
where Avg.~Time denotes the average
time required to analyze a single benchmark 
(i.e., one target function).
Since a single execution of \badger targets only a fixed input size, we run
\badger with input sizes $N$ ranging from 2 to 6 for each benchmark; at least
five data points are required to reliably distinguish among \cLowFull,
\cPolyFull, and \cExpFull time complexity classes.
In contrast, \spfWca is executed using its default configuration, except for the
TSP benchmark, where the maximum input size is set to $N=7$, as $N=8$ cannot be
completed within the 72-hour time limit.
As shown in the table, \proj achieves a substantially shorter average
per-benchmark analysis time, completing the analysis \speedupSpf{} faster than
\spfWca and \speedupBadger{} faster than \badger.

\subsection{Sensitivity Analysis}
\label{sec:sensitivity-analysis}
\newcommand{\sensDefaultFpr}{12.7}
\newcommand{\sensDefaultFnr}{29.0}
\newcommand{\sensDefaultPrec}{64.5}
\newcommand{\sensDefaultRecall}{71.0}
\newcommand{\sensDefaultFone}{67.6}
\newcommand{\sensNoStrategyFpr}{11.8}
\newcommand{\sensNoStrategyFnr}{46.4}
\newcommand{\sensNoStrategyPrec}{59.7}
\newcommand{\sensNoStrategyRecall}{53.6}
\newcommand{\sensNoStrategyFone}{56.5}
\newcommand{\sensNoCtxFpr}{14.2}
\newcommand{\sensNoCtxFnr}{47.8}
\newcommand{\sensNoCtxPrec}{54.5}
\newcommand{\sensNoCtxRecall}{52.2}
\newcommand{\sensNoCtxFone}{53.3}
\newcommand{\sensNoDynFpr}{3.3}
\newcommand{\sensNoDynFnr}{92.8}
\newcommand{\sensNoDynPrec}{41.7}
\newcommand{\sensNoDynRecall}{7.2}
\newcommand{\sensNoDynFone}{12.3}
\newcommand{\sensGptFourOneFpr}{17.5}
\newcommand{\sensGptFourOneFnr}{26.1}
\newcommand{\sensGptFourOnePrec}{58.0}
\newcommand{\sensGptFourOneRecall}{73.9}
\newcommand{\sensGptFourOneFone}{65.0}
\newcommand{\sensDeepseekFpr}{16.5}
\newcommand{\sensDeepseekFnr}{34.8}
\newcommand{\sensDeepseekPrec}{56.2}
\newcommand{\sensDeepseekRecall}{65.2}
\newcommand{\sensDeepseekFone}{60.4}
\newcommand{\sensHypofuzzFpr}{0.0}
\newcommand{\sensHypofuzzFnr}{92.8}
\newcommand{\sensHypofuzzPrec}{100.0}
\newcommand{\sensHypofuzzRecall}{7.2}
\newcommand{\sensHypofuzzFone}{13.5}
In this section, we present a sensitivity analysis of \proj, focusing on four factors: 
(a) Context Recovery, 
(b) Multiple input-generation strategies,
(c) Dynamic Validation, and (d) \llm model selection.
Since Stage~1 Static Analysis only performs high-density candidate selection without classification, 
we evaluate it separately from downstream accuracy.
For all other configurations, a function is considered an \acv if it is classified as \cPolyFull or \cExpFull. 
We evaluate each setting against the ground truth defined 
in Section~\ref{sec:experimental-setup}, and report FPR, FNR, precision, recall, and F1 score.
Table~\ref{tab:sensitivity-summary} summarizes configurations that
disable key components or replace the underlying \llm backend.

\myparagraph{Static Analysis}
Static Analysis performs lightweight static analysis for high-density candidate
selection, quickly identifying target functions that may exhibit \acvs.
In the \texttt{email} module (529 functions), 
Stage~1 selects \numCPythonEmailFuncs targets and prunes the remaining 248.
Among the pruned functions, only two contain previously unseen dominant \acv
sinks.
Thus, Stage~1 misses only two previously unseen dominant sinks while removing
46.9\% of functions from consideration.

\myparagraph{Effectiveness of Context Recovery}
The contribution of Context Recovery is evaluated using a variant w/o Context.
In this setting, \proj generates \Candidates using only the target function
and omits recovered context.
This substantially increases the FNR, because unresolved symbols can determine
the conditions and state needed to trigger high-cost paths.
With Context Recovery enabled, \proj reconstructs these dependencies and
detects additional \acvs that cannot be exposed from incomplete contexts.

\myparagraph{Effectiveness of Multiple Input-Generation Strategies}
Compared with the default setting, \emph{w/o Strat.} reduces recall from
\sensDefaultRecall\% to \sensNoStrategyRecall\% and F1 from
\sensDefaultFone\% to \sensNoStrategyFone\%, while increasing FNR from
\sensDefaultFnr\% to \sensNoStrategyFnr\%.
Its FPR remains nearly unchanged
(\sensDefaultFpr\% versus \sensNoStrategyFpr\%).
This indicates that using only a single generator mainly causes \proj to miss
more true \acvs, rather than substantially reducing false positives.
These results show that multiple strategies expose worst-case behavior more
effectively than a single generator.

\myparagraph{Impact of Dynamic Validation}
The contribution of dynamic validation is evaluated using a variant w/o
dynamic validation.
In the w/o Dyn.\ setting, \proj classifies complexity from the recovered
context and generated \Candidate without runtime measurements.
This conservative setting lowers the FPR from \sensDefaultFpr\% to
\sensNoDynFpr\%, but sharply increases the FNR from \sensDefaultFnr\% to
\sensNoDynFnr\%, causing most true \acvs to be missed.
With dynamic validation enabled, \proj grounds complexity inference in
execution-time measurements across input sizes, substantially improving recall
and F1.

\myparagraph{Sensitivity to \llm Choice}
The robustness of \proj to different \llm backends is evaluated by running
\proj with \defaultLLMModel~\cite{deepseekai2026deepseekv4},
\gptFourOne~\cite{openai2025gpt41}, and
\deepseek~\cite{deepseekai2025deepseekv32} under identical experimental
settings.
Across these \llms, \proj achieves broadly comparable performance, with only
moderate variations in FPR, FNR, precision, recall, and F1.
This suggests that \proj does not rely on a single proprietary model and that
its measurement-based validation provides stable performance across different
\llm backends.

\myparagraph{Comparison with Cost-Guided Fuzzing}
\hypofuzz~\cite{hypofuzz}, a feedback-directed Python fuzzer, is evaluated
on \emailBench using runtime as the feedback objective.
We use runtime as the feedback objective and classify a target as
an \acv when replay validation over the best discovered input family shows
\cPolyFull or \cExpFull growth.  Making this baseline work required
non-trivial adaptation: we had to construct specific generation
strategies, connect them to recovered executable harnesses, and replay the best
input families for complexity classification.  Even after this manual
adaptation, \hypofuzz produces no false positives but finds only 5 of
\numCPythonEmailACV
\acv-positive targets in the full benchmark, yielding a high FNR.  This result
indicates that cost-guided fuzzing can discover some worst-case inputs, but its
coverage on parsing-heavy \api surfaces depends heavily on time-consuming
domain-specific strategy engineering.

\myparagraph{Cost and Scalability}
\llm usage in \proj is localized to worst-case input construction.
Static Analysis, Context Recovery, and Dynamic Validation do not require
\llm calls.
On \emailBench, using \defaultLLMModel, each target function consumes
32K tokens on average, including 23K input tokens and 9K output tokens.
This corresponds to an average monetary cost of approximately \$0.006 per
target.

\begin{table}[t]
\centering
\caption{Sensitivity analysis of major system components, LLM backends, and a cost-guided fuzzing baseline.}
\label{tab:sensitivity-summary}
\scriptsize
\setlength{\tabcolsep}{2pt}
\renewcommand{\arraystretch}{1.03}
\resizebox{\columnwidth}{!}{%
\begin{tabular}{@{}l c c c l c c c c c@{}}
\toprule
Setting &
\multicolumn{3}{c}{Components} &
LLM &
\multicolumn{5}{c}{Metrics (\%)} \\
\cmidrule(lr){2-4}\cmidrule(lr){6-10}
&
Ctx. &
Strat. &
Dyn. &
&
FPR$\downarrow$ &
FNR$\downarrow$ &
Prec.$\uparrow$ &
Recall$\uparrow$ &
F1$\uparrow$ \\
\midrule
\textbf{Default}
& \checkmark & \checkmark & \checkmark & \defaultLLMModelShort
& \textbf{\sensDefaultFpr} & \textbf{\sensDefaultFnr} & \textbf{\sensDefaultPrec}
& \textbf{\sensDefaultRecall} & \textbf{\sensDefaultFone} \\

w/o Ctx.
& $\times$ & \checkmark & \checkmark & \defaultLLMModelShort
& \sensNoCtxFpr & \sensNoCtxFnr & \sensNoCtxPrec
& \sensNoCtxRecall & \sensNoCtxFone \\

w/o Strat.
& \checkmark & $\times$ & \checkmark & \defaultLLMModelShort
& \sensNoStrategyFpr & \sensNoStrategyFnr & \sensNoStrategyPrec
& \sensNoStrategyRecall & \sensNoStrategyFone \\

w/o Dyn.
& \checkmark & \checkmark & $\times$ & \defaultLLMModelShort
& \sensNoDynFpr & \sensNoDynFnr & \sensNoDynPrec
& \sensNoDynRecall & \sensNoDynFone \\

GPT-4.1
& \checkmark & \checkmark & \checkmark & GPT-4.1
& \sensGptFourOneFpr & \sensGptFourOneFnr & \sensGptFourOnePrec
& \sensGptFourOneRecall & \sensGptFourOneFone \\

DS-V3.2
& \checkmark & \checkmark & \checkmark & DS-V3.2
& \sensDeepseekFpr & \sensDeepseekFnr & \sensDeepseekPrec
& \sensDeepseekRecall & \sensDeepseekFone \\
\midrule
\multicolumn{10}{@{}l}{\emph{Cost-guided fuzzing baseline on \emailBench}} \\
HypoFuzz
& -- & -- & -- & --
& \sensHypofuzzFpr & \sensHypofuzzFnr & \sensHypofuzzPrec
& \sensHypofuzzRecall & \sensHypofuzzFone \\
\bottomrule
\end{tabular}
}
\end{table}

\subsection{RQ3: Real-World Detection}

\begin{table}[t]
\centering
\caption{Real-world projects evaluated by \proj and status of real-world \acv reports.}
\label{tab:real-world-project-versions}
\label{tab:bug-count}
\label{tab:real-world-effectiveness}
\scriptsize
\setlength{\tabcolsep}{1.5pt}
\begin{tabular*}{\linewidth}{@{\extracolsep{\fill}} l l r r r r @{}}
\toprule
Language & Project & Found & Confirmed & In Prog. & Fixed \\
\midrule
Python & \cpython (\cpythonCommit) & \numCpythonFound & \numCpythonConfirmed & \numCpythonFixing & \numCpythonFixed \\
Java   & \jdk (\jdkCommit)         & \numJdkFound     & \numJdkConfirmed     & \numJdkFixing     & \numJdkFixed \\
Zig    & \zig (\zigCommit)         & \numZigFound     & \numZigConfirmed     & \numZigFixing     & \numZigFixed \\
Rust   & \rustc (\rustcCommit)     & \numRustcFound   & \numRustcConfirmed   & \numRustcFixing   & \numRustcFixed \\
Python & \vllm (\vllmCommit)       & \numVLLMFound    & \numVLLMConfirmed    & \numVLLMFixing    & \numVLLMFixed \\
\midrule
\multicolumn{2}{l}{Total} & \numRealWorldFoundTotal & \numRealWorldConfirmedTotal & \numRealWorldFixingTotal & \numRealWorldFixedTotal \\
\bottomrule
\end{tabular*}
\end{table}

To answer RQ3, we apply \proj to the five real-world projects listed in
Table~\ref{tab:bug-count}: \cpython, \jdk, \zig, \rustc, and \vllm.
These projects cover language runtimes and infrastructure software written
in Python, Java, Zig, and Rust.
For each language, we implemented only the language-specific parts of the pipeline:
parsing and symbol extraction. 
This required \pythonAdapterLOC LOC for Python, \javaAdapterLOC LOC for Java,
\zigAdapterLOC LOC for Zig, and \rustAdapterLOC LOC for Rust; the Python
implementation was reused unchanged for both \cpython and \vllm.

Table~\ref{tab:bug-count} reports the \acv cases found by \proj and the
subsequent developer response.
In the table, 
\emph{Found} denotes cases reported by \proj after manual triage.
\emph{Confirmed} denotes cases acknowledged by maintainers as valid issues.
\emph{In Progress} and \emph{Fixed} summarize subsequent developer
responses. 
Unconfirmed cases are not treated as false positives.
They remain pending maintainer confirmation or disclosure discussion.

Across these projects, \proj finds \numRealWorldFoundTotal real-world
\acv cases, of which \numRealWorldConfirmedTotal have been confirmed by
maintainers.
Among the confirmed cases, \numRealWorldFixedTotal have been fixed and
\numRealWorldFixingTotal are in progress.
The largest numbers of findings come from \cpython and \jdk, where
\proj analyzes thousands of functions in mature language runtimes.
The findings in \zig, \rustc, and \vllm further demonstrate the applicability
of the pipeline across different language implementations and infrastructure
systems.

\section{Real-World Findings and Case Studies}
\label{sec:real-world}

Building on the real-world detection results, we further analyze
where the findings occur and what implementation patterns they reveal.
This section first examines their module-level distribution, then summarizes
common root causes and representative case studies.

\subsection{Distribution and Root Causes}

We further examine where the real-world \acv findings occur within
\cpython and \jdk.
These two projects contain enough findings to reveal module- and
package-level patterns.

\begin{figure}[t]
\centering
\footnotesize
\setlength{\tabcolsep}{3pt}
\renewcommand{\arraystretch}{1.05}

\newcommand{\numCPythonTopFiveModuleACVRaw}{33}
\newcommand{\numJdkTopFivePackageACVRaw}{20}

\newcommand{\spark}[3]{%
  \pgfmathsetmacro{\p}{#1/#2}%
  \pgfmathsetlengthmacro{\W}{1.05cm}%
  \pgfmathsetlengthmacro{\w}{\p*\W}%
  \begin{tikzpicture}[baseline=-0.6ex]
    \fill[black!12] (0,0) rectangle (\W,0.95ex);
    \fill[#3]      (0,0) rectangle (\w,0.95ex);
  \end{tikzpicture}%
}
\newcommand{\pct}[2]{%
  \pgfmathparse{100*(#1)/(#2)}%
  \pgfmathprintnumber[fixed,precision=0]{\pgfmathresult}%
}

\begin{tabular}{@{}l c r @{\hspace{6pt}} l c r@{}}
\toprule
\multicolumn{3}{c}{\textbf{\cpython}} & \multicolumn{3}{c}{\textbf{\jdk}} \\
\cmidrule(lr){1-3}\cmidrule(lr){4-6}
\textit{Module} & & \textit{\% ACVs} & \textit{Package} & & \textit{\% ACVs} \\
\midrule
\texttt{argparse}     & \spark{8}{\numCPythonTopFiveModuleACVRaw}{orange!70}  & \pct{8}{\numCPythonTopFiveModuleACVRaw} &
\texttt{sun.security} & \spark{6}{\numJdkTopFivePackageACVRaw}{blue!60}     & \pct{6}{\numJdkTopFivePackageACVRaw}  \\
\texttt{idlelib}      & \spark{8}{\numCPythonTopFiveModuleACVRaw}{orange!70}  & \pct{8}{\numCPythonTopFiveModuleACVRaw} &
\texttt{java.internal}& \spark{5}{\numJdkTopFivePackageACVRaw}{blue!60}     & \pct{5}{\numJdkTopFivePackageACVRaw}  \\
\texttt{xml}          & \spark{8}{\numCPythonTopFiveModuleACVRaw}{orange!70}  & \pct{8}{\numCPythonTopFiveModuleACVRaw} &
\texttt{jdk.util}     & \spark{3}{\numJdkTopFivePackageACVRaw}{blue!60}     & \pct{3}{\numJdkTopFivePackageACVRaw}  \\
\texttt{\_pyrepl}     & \spark{6}{\numCPythonTopFiveModuleACVRaw}{orange!70}  & \pct{6}{\numCPythonTopFiveModuleACVRaw} &
\texttt{sun.net}      & \spark{3}{\numJdkTopFivePackageACVRaw}{blue!60}     & \pct{3}{\numJdkTopFivePackageACVRaw}  \\
\texttt{inspect}      & \spark{3}{\numCPythonTopFiveModuleACVRaw}{orange!70}  & \pct{3}{\numCPythonTopFiveModuleACVRaw} &
\texttt{java.net}     & \spark{3}{\numJdkTopFivePackageACVRaw}{blue!60}     & \pct{3}{\numJdkTopFivePackageACVRaw}  \\
\bottomrule
\end{tabular}

\caption{
  Top-5 packages/modules with the largest shares of identified \acvs in \cpython and \jdk.
}
\label{fig:acv-top5-modules}
\end{figure}

Figure~\ref{fig:acv-top5-modules} reports the largest
\acv-containing modules and packages in \cpython and \jdk.
For \cpython, the dominant \texttt{email} module is analyzed
separately in Section~\ref{par:email}; the figure shows the
largest remaining modules.
Overall, the findings are concentrated in parsing- and text-processing
modules.
\texttt{argparse}, \texttt{idlelib}, and \texttt{xml} account for the
largest shares, followed by \texttt{\_pyrepl} and \texttt{inspect}.
These modules frequently parse structured text, manipulate strings, or
incrementally build outputs, which can amplify copying and scanning costs.
In the \jdk, findings are less concentrated and mainly appear in
security, utility, and networking packages.
Overall, the findings concentrate in modules that process structured or
text-heavy inputs.

To understand this concentration, we manually inspect the confirmed
polynomial-time findings and identify several recurring implementation
patterns.
The dominant pattern is \emph{copy amplification from immutability}.
This pattern appears when parsers repeatedly slice input buffers
(\eg, \texttt{s = s[i:]}) or when builders repeatedly append to a growing
output (\eg, \texttt{out += chunk}).
Although each operation is locally linear, placing it inside an
input-dependent loop yields quadratic behavior.

A second pattern is \emph{linear-cost list operations inside loops}.
Examples include queue-like use of \texttt{pop(0)} and repeated scans over
growing lists.
We also observe ordering costs and multi-parameter scaling, where both the
number of items and the size of each item grow with the input.
These patterns explain why parsing-heavy and text-processing modules appear
prominently in the distribution.

\subsection{Case Study: \cpython}
\myparagraph{Long-lived \texttt{singledispatch} case}
\proj detects a long-lived exponential case in Python's
\texttt{singledispatch} implementation in \texttt{functools}.
Its internal \texttt{\_c3\_mro} routine was introduced in 2013
(commit \texttt{3720c77e}) as part of PEP~443~\cite{pep443}.
The routine implements C3 superclass linearization
\cite{DBLP:conf/oopsla/BarrettCHMPW96,simionato2003mro}
in pure Python.
However, it recursively recomputes sub-results for each base class
without memoization.

Figure~\ref{fig:fib-inheritance} shows the adversarial inheritance pattern
used to trigger this behavior.
Each class inherits from the previous two classes, causing repeated
recursive computation of overlapping superclass linearizations.
As shown in Figure~\ref{fig:c3-mro-timing}, the vulnerable implementation
grows exponentially with inheritance depth, while a memoized implementation
removes the blowup.
This case shows that \acvs can remain latent for years in mature
standard-library code, even when the underlying fix is local.

\begin{figure}[h]
	\centering
	\begin{tikzpicture}[node distance=0.8cm]
		\node (c0) {$C_0$};
		\node (c1) [right=of c0] {$C_1$};
		\node (c2) [right=of c1] {$C_2$};
		\node (c3) [right=of c2] {$C_3$};
		\node (c4) [right=of c3] {$C_4$};
		\node (dots) [right=of c4] {$\cdots$};

		\draw[->] (c1) -- (c0);
		\draw[->] (c2) -- (c1);
		\draw[->] (c2) to[bend right=30] (c0);
		\draw[->] (c3) -- (c2);
		\draw[->] (c3) to[bend right=30] (c1);
		\draw[->] (c4) -- (c3);
		\draw[->] (c4) to[bend right=30] (c2);
	\end{tikzpicture}
	\caption{Fibonacci-style class inheritance pattern: each class $C_i$ inherits from $C_{i-1}$ and $C_{i-2}$.}
	\label{fig:fib-inheritance}
\end{figure}
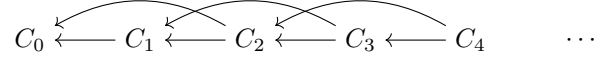
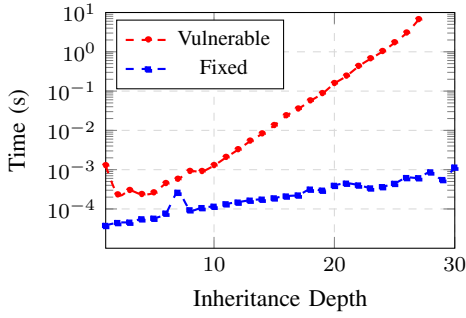
\begin{figure}[t]
	\centering
	\begin{tikzpicture}
		\begin{semilogyaxis}[
				width=0.7\columnwidth,
				height=4.7cm,
				xlabel={Inheritance Depth},
				ylabel={Time (s)},
				xmin=1, xmax=30,
				ymin=0.00001, ymax=10,
				ytick={0.0001, 0.001, 0.01, 0.1, 1, 10},
				legend pos=north west,
				legend style={font=\footnotesize},
				grid=major,
				grid style={dashed, gray!30},
				tick label style={font=\footnotesize},
				label style={font=\small},
			]
			\addplot[
				color=red,
				mark=*,
				mark size=1pt,
				dashed,
				smooth,
				thick
			] coordinates {
					(1, 0.001262) (2, 0.000228) (3, 0.000295) (4, 0.000233)
					(5, 0.000257) (6, 0.000443) (7, 0.000569) (8, 0.000901)
					(9, 0.000896) (10, 0.001272) (11, 0.002050) (12, 0.003220)
					(13, 0.005325) (14, 0.008148) (15, 0.013261) (16, 0.023539)
					(17, 0.034974) (18, 0.056331) (19, 0.086255) (20, 0.155637)
					(21, 0.240032) (22, 0.424371) (23, 0.664155) (24, 1.016321)
					(25, 1.698718) (26, 3.008221) (27, 6.578612)
				};
			\addlegendentry{Vulnerable}

			\addplot[
				color=blue,
				mark=square*,
				mark size=1pt,
				dashed,
				smooth,
				thick
			] coordinates {
					(1, 0.000038) (2, 0.000045) (3, 0.000046) (4, 0.000056)
					(5, 0.000058) (6, 0.000078) (7, 0.000265) (8, 0.000095)
					(9, 0.000108) (10, 0.000117) (11, 0.000135) (12, 0.000149)
					(13, 0.000167) (14, 0.000177) (15, 0.000191) (16, 0.000213)
					(17, 0.000226) (18, 0.000319) (19, 0.000300) (20, 0.000398)
					(21, 0.000452) (22, 0.000406) (23, 0.000342) (24, 0.000369)
					(25, 0.000447) (26, 0.000628) (27, 0.000622) (28, 0.000875)
					(29, 0.000559) (30, 0.001159)
				};
			\addlegendentry{Fixed}

		\end{semilogyaxis}
	\end{tikzpicture}
	\caption{Execution time: vulnerable vs.\ fixed \texttt{\_c3\_mro}.}
	\label{fig:c3-mro-timing}
\end{figure}

\myparagraph{Shared \texttt{email} parser}
\label{par:email}
The \texttt{message.py} module in \cpython's \texttt{email} package
implements core MIME-message logic, including the public
\texttt{Message} and \texttt{EmailMessage} abstractions.
It is used by mail-related software and application frameworks that process
MIME-formatted data, such as Django's email subsystem, Flask-Mail, and GNU
Mailman~\cite{django,flask-mail,gnu-mailman}.
These APIs often operate on partially user-controlled inputs, including
message headers and multipart parameters.

\begin{figure}[t]
\centering
\includegraphics[width=0.80\columnwidth]{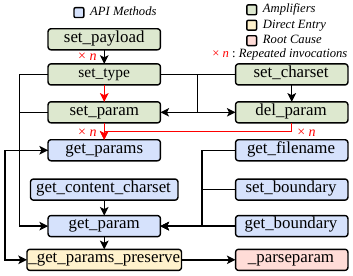}
\caption{Structural view of exposure, trigger, and root cause for the \acv in
\texttt{email/message.py}.
}
\label{fig:email-layered-acv}
\end{figure}

Figure~\ref{fig:email-layered-acv} shows how several public APIs share the
same underlying complexity bottleneck.
Parameter-related methods eventually call
\texttt{\_get\_params\_preserve}. This helper reparses the header value
and invokes the quadratic-time parser \texttt{\_parseparam}.
Thus, even simple parameter queries can trigger the same worst-case parser.
Mutating APIs further reconstruct the header, and higher-level routines may
repeat these operations in loops.
As a result, multiple findings in \texttt{email/message.py} are different
entry points to a shared quadratic parser rather than independent bugs.

\subsection{Case Study: \zig}
\myparagraph{\zig compiler frontend case}
\proj also detects a compile-time \acv in \zig's compiler frontend.
The trigger is a valid integer \texttt{switch} statement with many prong
items.
During semantic analysis, \zig records previously seen integer ranges in a
\texttt{RangeSet}; the original implementation linearly scans existing
ranges on each insertion to detect overlaps.
This makes switch validation quadratic in the number of prongs.
When the number of switch branches becomes large, compilation time increases significantly and can become prohibitively slow.
The issue applies to all optimization levels and targets because semantic
analysis runs before backend code generation.
The merged fix keeps the range set sorted and uses binary search to check
overlaps, reducing the comparison cost to \(O(n\log n)\).
This case illustrates a compiler front end \acv: an adversarial source
program can cause large compile-time slowdowns.

\section{Discussion}

\subsection{Implications for Library Maintenance}

Our real-world findings suggest two implications for maintaining mature
libraries and language runtimes.
First, long-lived parsing and text-processing code should be revisited even
when its functional behavior remains stable.
In \cpython's \texttt{email} module, some vulnerable code dates back to
2007, and incremental maintenance allowed inefficient helper patterns to
persist across many public entry points.
This suggests that complexity-oriented testing can complement functional
regression testing for mature library code.
Second, implementation-specific optimizations should not be treated as a
substitute for robust algorithmic design.
For example, \cpython can optimize some repeated string concatenation
patterns in loops~\cite{Kuchling2005WhatsNewPython24Optimizations,
BolzTereick2023RepeatedStringConcatenationQuadratic}, but the optimization
applies only under specific conditions and may not hold across runtimes.
Using index-based parsing, explicit buffers, or builder-style accumulation
provides more portable performance behavior.

\subsection{Limitations}

\proj is an empirical detection system rather than a complete oracle for all
possible \acvs.
Although our evaluation covers multiple ecosystems and languages, extending
\proj to additional languages or domains may require engineering effort for
integration and adaptation to new environments.
\proj validates findings through executable \Candidates and measured runtime
growth, providing evidence for developer triage but not replacing maintainer
assessment.
In particular, whether a detected complexity issue should be treated as a
security vulnerability or a performance bug can depend on the specific usage
context and deployment conditions.

\section{Related Work}
\noindent{\textbf{Symbolic Execution and Fuzzing.}}
Crosby and Wallach~\cite{DBLP:conf/uss/CrosbyW03} introduced algorithmic
complexity attacks, showing that adversarial inputs can induce worst-case
execution and denial of service.
Subsequent work explored automated
detection using fuzzing and symbolic execution.
Resource-guided fuzzers
such as SlowFuzz~\cite{DBLP:conf/ccs/PetsiosZKJ17} and
HotFuzz~\cite{DBLP:conf/ndss/BlairMAW0KE20} employ evolutionary search to
maximize resource usage, while Singularity~\cite{DBLP:conf/sigsoft/WeiCFFD18}
frames complexity testing as synthesis over input generators.
Symbolic
approaches, including WISE~\cite{DBLP:conf/icse/BurnimJS09} and
SPF-WCA~\cite{DBLP:conf/icst/LuckowKP17}, provide precise path-level reasoning
but suffer from path explosion, with extensions such as
XSTRESSOR~\cite{DBLP:conf/icst/SaumyaK0B19} mitigating this limitation.
Hybrid systems such as Badger~\cite{DBLP:conf/issta/NollerKP18},
ACQUIRER~\cite{DBLP:conf/ccs/Liu022}, and
DISCOVER~\cite{DBLP:conf/sigsoft/AwadhutkarSHK19} combine static analysis
with guided dynamic exploration, sometimes incorporating human guidance.
However, these approaches often have difficulty automatically constructing
calling contexts in complex programs.
In contrast, \proj combines context recovery with 
\llm-assisted worst-case input construction and measurement-based
validation to generate executable evidence at scale.

\noindent{\textbf{LLM-assisted Program Analysis.}} 
With the advent of \llms, several efforts have examined their use for
vulnerability detection and repair~\cite{DBLP:journals/tosem/ZhouCSL25,
DBLP:conf/icst/Khare0LSAN25,DBLP:journals/corr/abs-2401-16185,
DBLP:conf/uss/0003LFLX0CWJW24}.
These studies further evaluate \llms across diverse vulnerability
classes and datasets, and explore their effectiveness in related
management tasks such as triaging, severity assessment, and patch
validation.
Our work differs in that we focus on algorithmic complexity
vulnerabilities rather than general vulnerability detection
or management tasks. 
We study how \llms can be used as a
supporting component in a language-extensible analysis
pipeline to reason about program behavior related to
time complexity.

\noindent{\textbf{\acv at Different Layers.}}
Prior work has studied algorithmic complexity vulnerabilities at the network,
protocol, and application layers~\cite{DBLP:conf/tacas/WustholzOHD17,DBLP:conf/critis/KambourakisMGG07,DBLP:conf/uss/AfekBS20},
focusing on attacks such as request flooding and resource exhaustion.
Related efforts on regular-expression DoS~\cite{DBLP:conf/sp/LiuZM21,DBLP:conf/sigsoft/AwadhutkarSHK19}
address pathological patterns that cause excessive matching overhead.
In contrast, \proj targets algorithmic inefficiencies in software components,
exposing complexity vulnerabilities that can trigger DoS even
without high network traffic.

\section{Conclusion}
Algorithmic complexity vulnerabilities remain difficult to detect because
their triggering conditions often depend on subtle interactions between program
structure, input dependencies, and opaque callee behavior.
In this work, we identify \shadowComplexity as a critical blind spot in \acv
detection, where hidden computational costs behind API calls can silently
amplify worst-case execution behavior.

We present \proj, an \llm-assisted framework for discovering
algorithmic complexity vulnerabilities through lightweight screening,
context recovery, and measurement-driven validation.
Applying \proj to \cpython, \jdk, \zig, \rustc, and \vllm uncovered
\numRealWorldFoundTotal previously unknown \acvs, of which
\numRealWorldConfirmedTotal have been confirmed.

\bibliographystyle{IEEEtran}
\bibliography{reference}

\end{document}